\documentclass[preprint]{aastex}

\slugcomment{\it to appear in Astrophys. J., Vol. 531}

\shorttitle{Polarization under Anisotropic Radiation}
\shortauthors{T. ONAKA}

\begin{document}


\title{POLARIZATION OF THERMAL EMISSION FROM ALIGNED DUST GRAINS 
UNDER AN ANISOTROPIC RADIATION FIELD}


\author{TAKASHI ONAKA}
\affil{Department of Astronomy, School of Science, University of Tokyo,
    Tokyo 113-0033, Japan}

\email{onaka@astron.s.u-tokyo.ac.jp}



\begin{abstract}
If aspherical dust grains are immersed 
in an anisotropic radiation field, 
their temperature 
depends on the cross-sections projected in the direction of the anisotropy.
It was shown that the temperature difference produces polarized 
thermal emission even
without alignment, if the observer looks at the grains from a direction
different from the anisotropic radiation.  When the dust grains are
aligned, the anisotropy in the 
radiation makes various effects on the polarization
of the thermal emission, depending on the relative angle between the
anisotropy and alignment directions.  If the both directions are parallel,
the anisotropy produces
a steep increase in the polarization degree at short wavelengths.
If they are perpendicular, the polarization reversal occurs at a 
wavelength shorter than the emission peak.  The effect of the anisotropic 
radiation will make a change of more than a few \%
in the polarization degree for short wavelengths and
the effect must be taken into account
in the interpretation of the polarization in the thermal
emission.  The anisotropy in the radiation field produces a strong
spectral dependence of the polarization degree and position angle, 
which is not seen under isotropic radiation.  The dependence
changes with the grain shape to a detectable level
and thus it will provide a new tool to 
investigate the shape of dust grains.
This paper presents examples of numerical calculations of the
effects and demonstrates the importance of anisotropic radiation field 
on the polarized thermal emission.
\end{abstract}



\section{INTRODUCTION}
The polarization observation in the far-infrared region is 
a quite efficient means for the investigation of 
magnetic fields in dense regions and the alignment mechanism
of dust grains \citep{hil88}.
A large number of polarization observations have been reported recently
for various interesting objects in the far-infrared and the submillimeter regions 
\citep[e.g., ][]
{hil95, nov97, sch98, dow97, dow98}.
\citet{hil99} have provided a summary of the latest observations
and discussed several factors which could
affect the observed polarization and the spectral dependence, such
as the different populations of grains with different temperatures, 
other than the magnetic field and grain alignment.  
In this paper, another mechanism, the anisotropy
of the incident radiation field, which is believed to
occur commonly in celestial objects, 
is shown also to affect the polarization of thermal emission.

If aspherical grains are immersed in an anisotropic radiation field,
the grains with larger cross-sections projected in the radiation direction 
absorb more photons and thus have higher temperatures than those with
smaller cross-sections.  The former grains emit
the radiation polarized in the direction perpendicular to the incident 
radiation field, while the radiation from the latter is polarized 
in the parallel direction,
if one observes them from a direction different from the radiation
anisotropy.  Therefore
owing to the difference in the temperature, the resultant total thermal
emission is polarized perpendicularly to the incident radiation. 

\citet[hereafter Paper I]{onaka95} has made numerical calculations of this
effect and shown that it produces polarization of
a detectable degree ($\sim$ 1\%) around 100\,$\mu$m
and a steep rise in the polarization degree at shorter
wavelengths even without any net alignment.  This paper presents
the numerical results of several simple examples 
when aligned dust grains are immersed in an unidirectional radiation
field and discusses the effects of anisotropic radiation on the polarized 
thermal emission from aligned dust grains.  
In \S 2, the basic assumptions made in the present
study are described, and the numerical results of several examples are
presented in \S 3.  In \S 4, the results are examined in relation
to the observations and the effects of anisotropic radiation on the
polarization of thermal emission are discussed.

\section{BASIC CONFIGURATION}

\begin{figure}[h]
\epsscale{0.8}
\plotone{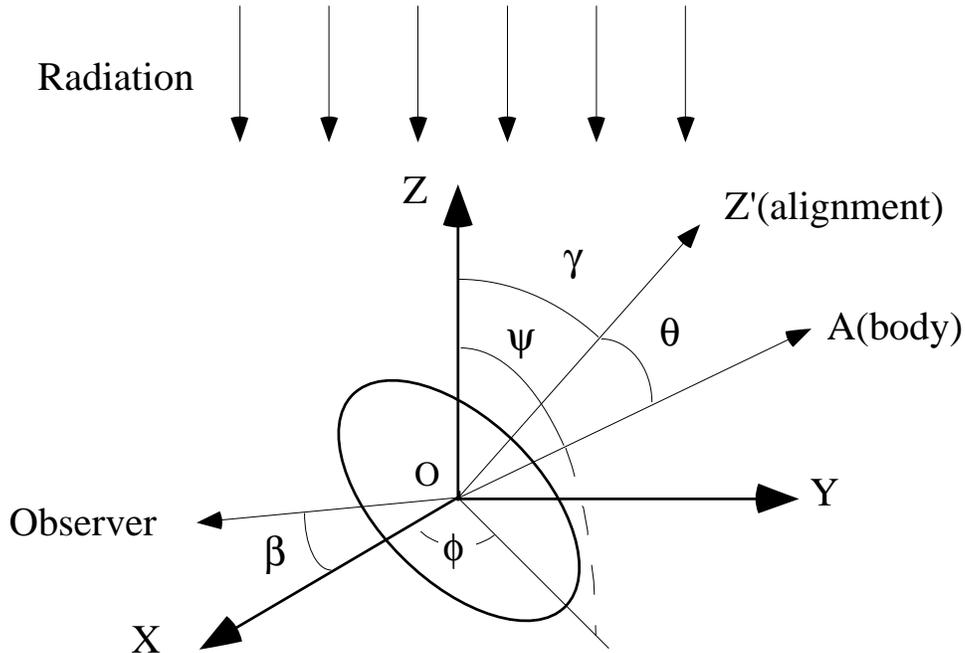}
\caption{A schematic diagram of the assumed configuration.  
In the numerical calculations presented in this paper $\beta$ is set as
zero.}
\end{figure}

Figure 1 depicts the geometry adopted in this paper:  The radiation
field is assumed to be unidirectional and coming from +Z direction;
The body axis of greatest inertia of the grain $A$ is tilted from the Z axis 
by $\psi$; The
angle between the alignment axis $Z'$ and the $Z$ axis
is denoted by $\gamma$ and the $Z'$ axis
is assumed to be in the Y-Z plane; The observer
is located in the X-Z plane with the angle $\beta$ from the X axis.  In the
following, $\beta$ is set as zero (the observer is on the
X axis) for simplicity.

Dust grains perform complex motions in the interstellar space; 
the grain axis $A$ makes a precession about
the angular momentum $J$ and $J$ makes a precession about the magnetic field
due to the magnetic moment of the grain \citep{laz97b}.  
The former angular velocity is of the order of the
grain angular velocity.  The Barnett effect, the inverse of the well-known 
Einstein-de Haas effect, produces the 
magnetic moment of the rotating grain, which is 
much lager than that acquired by a charged
rotating grain \citep{dol76}.  
The period of the precession due to the Barnett effect magnetic moment 
is of order of $10^6$~s \citep{laz94}.  It is still much longer than
the mean duration time of photon absorption in the interstellar radiation
field (ISRF) of $1-10$~s for $0.01 - 0.1$\,$\mu$m-sized grains \citep{pur76}.
Therefore, the angular momentum direction can be assumed to be fixed
in the reference frame as long as the temperature of the grain is considered.

Dust grains in the interstellar space are thought to rotate suprathermally 
due to the irregularities in the shape or surface characteristics
\citep{pur79, dra96}.  The rotation time scale is very short, typically
$10^{-7}$~to $10^{-9}$~s, depending on the grain size, and on the mechanism and
efficiency of suprathermal rotation under the 
physical conditions in question.  
Owing to
the internal energy dissipations, such as the anelasticity or the Barnett
effect relaxation, 
the rotation axis aligns with the body axis of greatest inertia
if the grain rotates suprathermally \citep{pur79}.  Most recently
\citet{lad99} have shown that nuclear spin relaxation within the grain
will be the most efficient process in the internal
energy dissipations.  On the other hand, \citet{laz94} pointed
out that the alignment is always not perfect and the deviation
should be correctly
taken into account when the internal temperature is comparable with the
grain kinetic temperature \citep{laz97a, lazrb97}.  The deviation is,
however, typically very small for suprathermally rotating 
grains \citep{laz94}.  
In this paper, the body axis $A$ is assumed to coincide with
the angular momentum $J$ for simplicity.

The alignment mechanism
of interstellar dust grains is still a hot issue.
Various mechanisms and their revisions have so far been proposed, including
paramagnetic relaxation of ordinary paramagnetic
grains {\citep{dav51} or grains with superparamagnetic inclusions
\citep{jon67, mathis86, mar95, goo95}, mechanical alignment
\citep{gold51, laz94}, alignment due to suprathermal rotation
\citep{pur79, laz95a, laz95b, lazdr97, lazdr99}, and  alignment by radiation
torques \citep{har70, dol76, dra96}.  The efficiency of each mechanism depends
on the physical conditions in question.  See \citet{lazgm97} for
a latest review and details of each mechanism.  
The time scale of paramagnetic
relaxation and mechanical alignment is, however, typically 
much longer than the photon
incident duration mentioned above.  The mean duration of the crossover
for suprathermally rotating grains must also be quite long if this
mechanism is effective in the grain alignment.
In this paper, I assume that the grain
has a fixed rotation axis, which coincides with the body
axis of greatest inertia, relative to the incident radiation
and thus the temperature of the grain is given by a function of $\psi$.

The alignment distribution of the dust grains around the alignment axis
depends on the mechanism of alignment.  In this paper, I assume the following
distribution function for simplicity:

\begin{equation}
f(\theta)~ d\cos\theta = \frac{0.5~q}{(\cos ^{2} \theta + q~\sin ^ {2} \theta
)^{3/2}}~ d\cos\theta,
\label{eq:equation1}
\end{equation}

\noindent where $q$ is a parameter of the alignment degree.  When no
alignment is established, $q$ is equal to unity.
For $q > 1$, the distribution peaks at $\theta = 0$, while
the perpendicular alignment is expressed by $q <1$.  In the
following, only the cases of $q>1$ are considered.

For the Davis-Greenstein alignment, the distribution is given by 
equation~(1) and $q$ is equal to $T_g / T_{av}$, where
$T_g$ is the gas temperature and $T_{av}$ is the weighted mean of 
gas and dust temperatures \citep{jon67}.  The distribution
with $q = T_g/T_{av}$ may be applicable only if grains are superparamagnetic
or ferromagnetic.  In the mechanical alignment
by supersonic flow the distribution is given by the same functional
form with $q = (1+g)/(1+g+s)$, where $g$ is a parameter of the grain
shape and $s$ is that of the flow anisotropy \citep{laz94}.
It can also be shown that equation~(1) holds 
for the alignment by suprathermal rotation with
$q$ being equal to exp$(2t_c/t_r)$, where $t_c$ and $t_r$ are the
time from the last crossover and the magnetic time scale, respectively, 
if the orientation after a crossover is completely
isotropic \citep{pur79}.  \citet{lazdr97, lazdr99} have shown that
the angular momentum before and after a crossover should be 
correlated.  This effect
may be taken into account by a slight modification in the parameter 
\citep{lazdr97}.  
When the coincidence of the body axis and the angular momentum axis
is not perfect, further modification of the distribution is
needed \citep{laz97a, lazrb97}.
The functional form
of equation~(1) may not always hold. It is adopted
here only to represent a simple alignment distribution.

According to \citet{lazdr97}, the grain alignment may be nearly perfect.
In the perfect alignment case, the present results may still be applicable
to the case that the alignment direction (the angle $\gamma$) changes along
the line of sight and the distribution of $\gamma$ is given by a
function similar to equation~(1) provided that the direction of the 
radiation field is fixed.

The measure of alignment $\sigma$ is defined by \citep{gre68}

\begin{equation}
\sigma = {3 \over 2} (<\cos^2\theta> - {1 \over 3}),
\end{equation}

\noindent where $<>$ denotes the distribution average.
Taking the ensemble average with equation~(1), one
obtains \citep{pur79}

\begin{equation}
\sigma = {3 \over 2} {q \over {q-1}} \{1- (q-1)^{-1/2} \tan^{-1}(q-1)^{1/2}\} - {1 \over 2}.
\end{equation}

Figure 2 plots the distributions given by 
equation~(1) with $q = 1.2$, 1.5, and 2.0.  These correspond to
$\sigma = 0.037$, 0.083, and 0.14, respectively.
It can be
seen that the
distribution is quite broad and the alignment is
far from being perfect for this range of $q$ .

\begin{figure}[h]
\epsscale{0.6}
\plotone{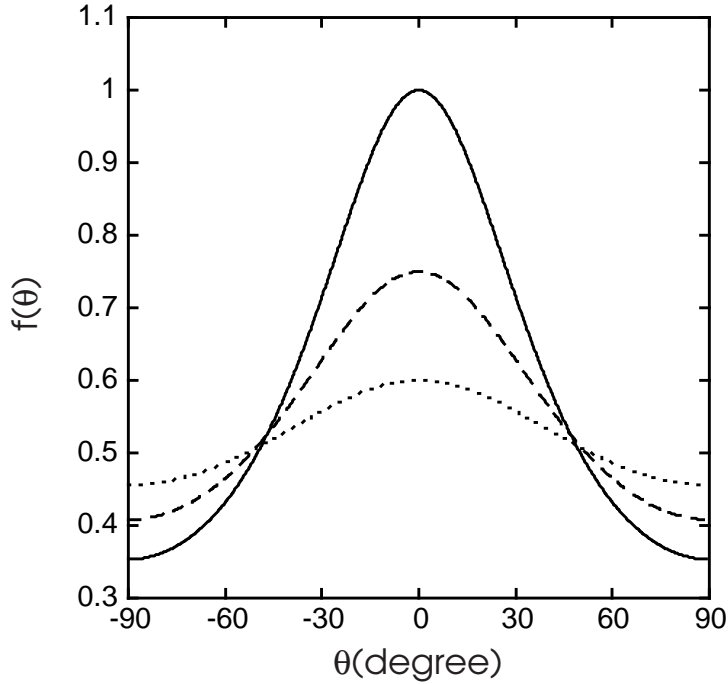}
\caption{The alignment distribution function $f(\theta)$
given by equation~(1).  The dotted line indicates the distribution
for $q=1.2$, the dashed line for $q=1.5$, and the solid line for
$q=2.0$.}
\end{figure}

The shape of the interstellar grains is not well known.  Several past investigations have suggested
that oblate grains fit the observations better than
prolate grains \citep{lee85, hen93, kim95, hil95b}.
\citet{hil95b} have indicated that oblate spheroidal grains with axial ratio
$b/a = 1.5$ fit the mid-infrared observations best, while \citet{kim95} have
suggested that flat oblate grains fit the visual 
polarization observations most satisfactorily.  
\citet{kim95} have also indicated that
dust grains smaller than 0.1\,$\mu$m are not well aligned.
On the other hand, 
alignment of silicate grains are indicated by the observed mid-infrared
band feature in the polarization spectrum \citep{ait88}.
 
In the following calculations, the grains are assumed to be oblate silicate
spheroids.  The calculations are made with the axial ratio $b/a = 1.5, 2.0$, 
and 3 to investigate
the dependence on the grain shape.  The particle size is fixed as
$a = 0.1$\,$\mu$m. 
The optical properties
are taken from those of astronomical silicate \citep{dra84} and the incident
radiation is adopted from the ISRF given by \citet{mat83}.  
The far-infrared emission
is assumed to be optically thin.  The absorption cross-sections of
spheroidal grains are calculated by the method of expansion series of
spheroidal functions developed by \citet{onaka80}.
Details of the formulation have been given in Paper I.

\section{NUMERICAL RESULTS}

\begin{figure}[h]
\epsscale{0.6}
\plotone{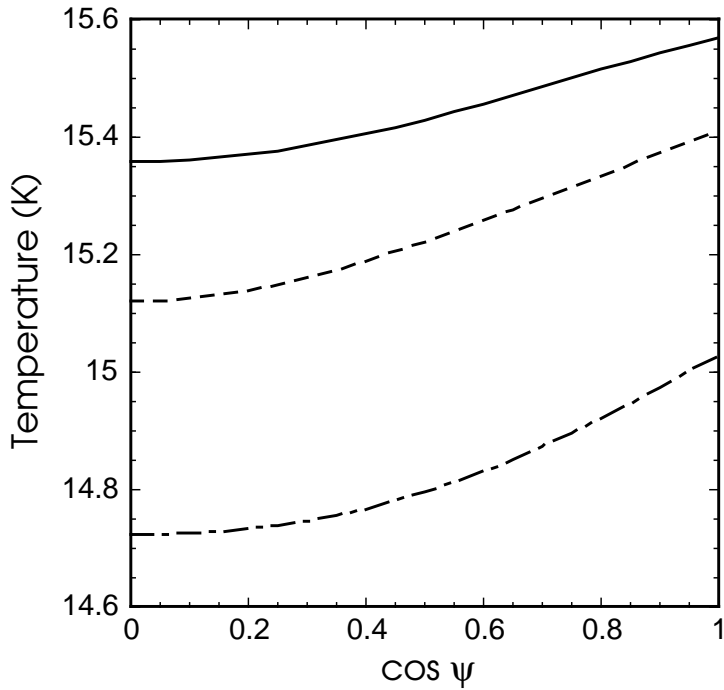}
\caption{The temperature of oblate silicate spheroid
grains of $a = 0.1$\,$\mu$m
in a unidirectional radiation field against the cosine
of tilt angle $\psi$.  The solid line indicates the grain 
with $b/a = 1.5$, the dashed line that with $b/a = 2$,
and the dot-dashed line that with $b/a = 3$.}
\end{figure}

The temperature of the grain is plotted against cos$\psi$ in Figure~3
for the parameters described in the previous section.  
The temperature
of the grain with small axial ratio is higher than that with large ratio
because the size of the major axis $a$ is fixed in the present calculation 
and thus the total volume is small for the grains with small axial ratio.
The difference in temperature between cos$\psi$=0 and 1
is quite small 
even for the oblate spheroid of axial ratio of 3. It
becomes smaller as the grain becomes less flattened.   
The effect is insignificant in the Rayleigh-Jeans tail of the
thermal emission, but becomes appreciable for shorter wavelengths
than the emission peak (see Paper I).  In this section, three
cases of the direction of alignment relative to the radiation direction
are investigated.

\begin{figure}[h]
\epsscale{1.0}
\plotone{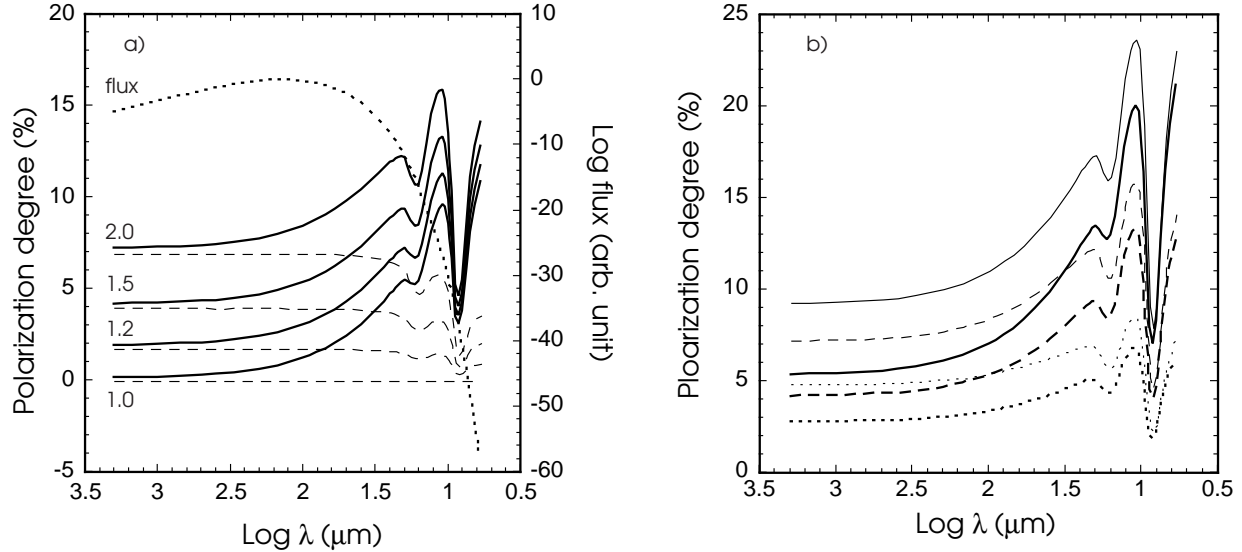}
\caption{a) Spectral dependence of the
polarization degree (left axis) for the alignment axis parallel with
the radiation direction ($\gamma = 0^\circ$) for the oblate spheroidal
grain of $b/a = 2$ together with the emission
spectrum (right axis in arbitrary units; dotted line labeled as flux). 
The dashed lines indicate the polarization degrees expected for the isotropic
radiation field, while the solid lines are those for a unidirectional
radiation field.  The numbers show the values of alignment parameter
$q$. b) The same as a) but for different grain shapes.
The solid lines indicate the results for $b/a =3$, the
dashed lines those for $b/a = 2.0$, and the dotted lines for
$b/a = 1.5$.  The thick lines are for $q = 1.5$ and the thin lines
are for $q = 2.0$.}
\end{figure}

Figure~4a shows the spectral dependence of the polarization degree 
(left axis; solid lines) of the oblate grain of $b/a=2$ for $\gamma = 
0^\circ$ together with the emission spectrum (right axis; dotted line
labeled as flux).  The results 
of isotropic radiation case are plotted together by dashed lines
for comparison.   As shown in Paper 
I, even without alignment ($q=1$), thermal emission of the dust grains in an 
anisotropic radiation field is polarized and the polarization degree increases 
steeply as the wavelength becomes shorter.  For the cases when the grains are aligned, 
the anisotropic radiation field increases the polarization degree
appreciably at short 
wavelengths compared to the isotropic radiation case.  Under the
isotropic radiation field, almost no spectral dependence is expected
for wavelengths longer than 30\,$\mu$m \citep{hil88}.  The strong
spectral dependence due to the anisotropic radiation field 
is quite remarkable.
The effect of anisotropic 
radiation field becomes more than a few \% in the polarization degree
particularly for wavelengths shorter than 
the thermal emission peak.  This can easily be understood from the nature 
of the mechanism.

\begin{figure}[h]
\epsscale{0.6}
\plotone{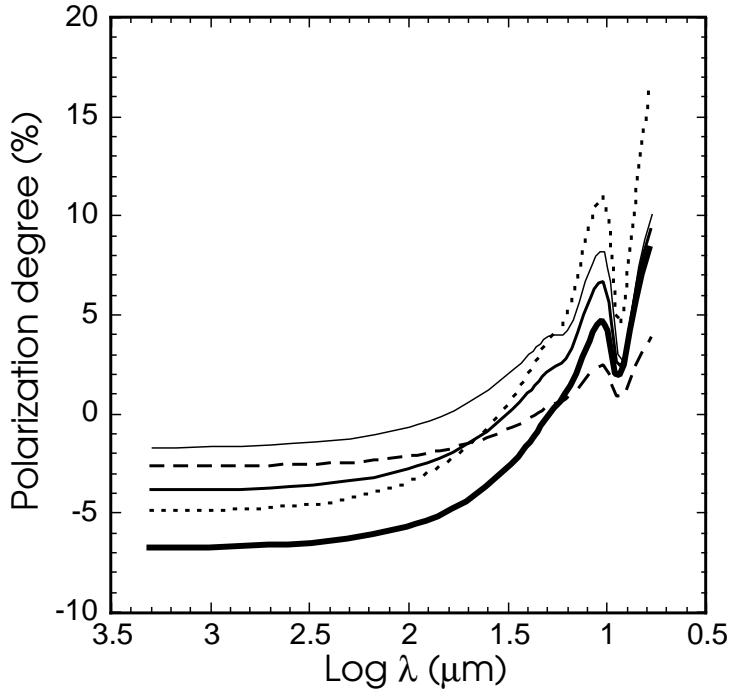}
\caption{Spectral dependence of the polarization degree
for the alingment axis perpendicular to the radiation direction
($\gamma = 90^\circ$).  The solid lines indicate the grain with
$b/a = 2.0$; the thin line for $q=1.2$, the medium-thickness line
for $q=1.5$, and the thick line for $q=2.0$.  The dashed line
shows the result for the grain of $b/a=1.5$ and $q=1.5$,
and the dotted line for $b/a = 3.0$ and $q=1.5$.}
\end{figure}

The spectral dependence of the polarization
degree changes with the grain shape
(Figure~4b).  Even when the polarization degree is similar at 
longer wavelengths (e.g., the grain of $b/a = 1.5$ with $q =2.0$
(thin dotted line) and that of $b/a = 3$ with $q=1.5$ (thick solid line))
the flatter grain shows a steeper spectral dependence.
This is because 
the effects of anisotropic radiation are larger for flatter grains
than less flatter grains due to the larger difference in the 
cross-sections.  Less alignment degree is required for 
flatter grains to produce a
given polarization degree at wavelengths longer than 100\,$\mu$m. 

When the alignment axis is perpendicular to the incident radiation direction
($\gamma = 90^\circ$), 
the  polarization degree of aligned grains first decreases
as the wavelength becomes shorter (Figure~5).  
Then the polarization reversal occurs at some wavelength where the effect
of anisotropic radiation starts dominating over the original alignment.  
In the calculation, the 
sign of the polarization degree is defined as positive if the emission is polarized 
in the Y axis.  The reversal wavelength depends on the alignment degree
and grain shape, and occurs at a wavelength shorter than the emission peak.
Figure~5 also 
indicates a similar trend of the spectral dependence with the grain shape
to that for $\gamma=0^\circ$.

\begin{figure}[h]
\epsscale{0.6}
\plotone{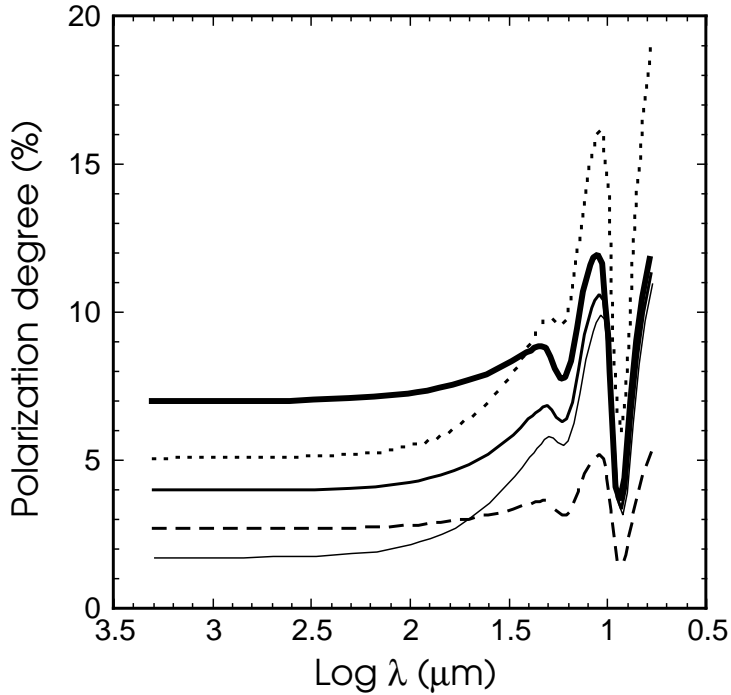}
\caption{Spectral dependence of the polarization degree
for $\gamma = 45^\circ$.  The symbols are the same as in Figure~5.
}
\end{figure}

If the alignment axis  is tilted against the radiation direction 
by an arbitrary angle, 
further complicated effects appear.  Figures~6 and 7 
plot the results for $\gamma 
= 45^\circ$.  The spectral dependence of the polarization degree is similar to that 
for $\gamma = 0^\circ$ or $90^\circ$, while the polarization angle also
varies with the wavelength 
(Figure~7).  The polarization angle is parallel with the alignment 
direction for long wavelengths ($45^\circ$), but it becomes 
perpendicular to the radiation 
direction as the wavelength becomes shorter.  The spectral dependence
of the polarization angle also varies with the grain shape and alignment degree.
For low alignment degree case ($q=1.2$, thin solid line) the deviation from
the original alignment direction can be seen even for wavelengths longer than
100\,$\mu$m.

\begin{figure}[h]
\epsscale{0.6}
\plotone{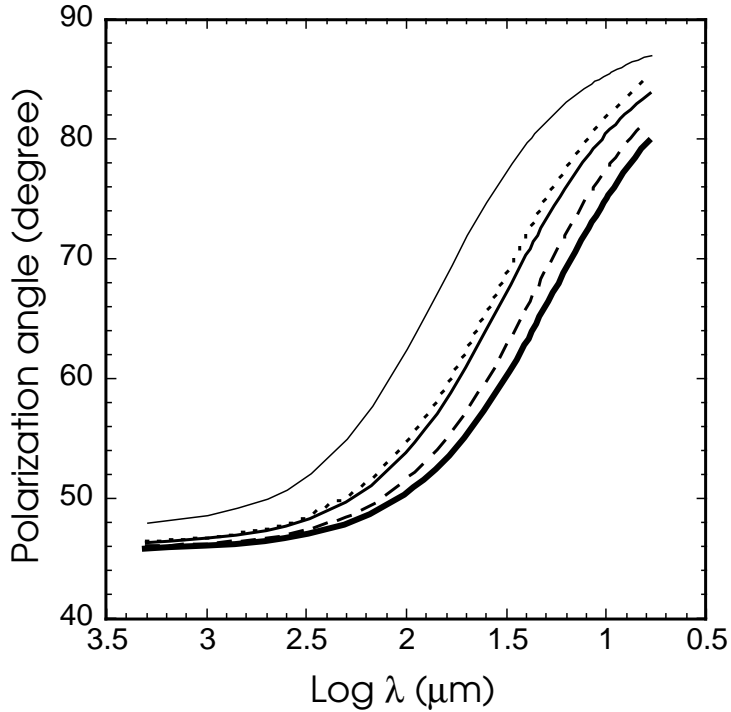}
\caption{Spectral dependence of the polarization
angle for $\gamma=45^\circ$. The symbols are the same as in Figure~5.
}
\end{figure}

\section{DISCUSSION}

\citet{hil99} have presented the latest results of far-infrared polarization 
measurements and shown that most of the observed polarization 
degrees in $60-100$\,$\mu$m
are in the range less than 10\%.  
The anisotropic radiation field produces very small differences
in temperature of the dust grains (Figure~3), but the effect in the 
polarization 
degree is still a few percent level for wavelengths shorter than 100\,$\mu$m,
being compatible with the observed range.  Thus the effect of anisotropic radiation
field cannot be neglected in the interpretation of the 
polarization observation of thermal emission.

\citet{hil99} have also shown the spectral dependence of the polarization degree
in the far-infrared region.
In the full samples presented, there is a general trend in the distribution of 
observed degrees of polarization that
the degree decreases with the wavelength, though this trend is seen in the 
combined data set obtained at different regions and/or objects.
In two particular objects the polarization degree either increases as the wavelength 
decreases (Orion core) or decreases sharply for short wavelengths
(M17).  In another case
\citet{dow97} showed that the polarization degree increases from submillimeters
to 60\,$\mu$m in Sgr B2 core and claimed that the polarization in the far-infrared in Sgr B2 core is due
to selective absorption.   In most other observed objects, however, 
the far-infrared optical depth
is not very large and the polarization occurs in emission.
\citet{hil99} attributed the observed spectral dependence to two populations of grains 
that differ in their polarization efficiencies and in their temperatures 
due to the difference in their optical properties.  

The mechanism proposed in the present paper also comes from the difference in the 
temperature of dust grains, but it originates from the anisotropy in the incident 
radiation field, which is not uncommon in most astronomical objects.  
It could make either positive or negative spectral dependence 
as being actually observed, 
depending on the configuration of the alignment and radiation field.  
The change in 
the polarization angle could also occur due to this mechanism.  
In the FIR polarization observations of Sgr B2, the observed change in the
polarization angle from the core to envelope was successfully explained
by the selective absorption in the core region, where the optical
depth is expected to be larger than unity \citep{dow97, nov97}.
The present mechanism, on the other hand, can produce polarization
angle variation even in optically-thin regions.
It should be noted that in the vicinity of the bright source the anisotropic
radiation could also be effective in the grain alignment \citep{dra96}.  

Because of the nature of the blackbody radiation, the effect due to the difference in
temperature appears most severely at short wavelengths.  The present study
has in fact shown that the polarization at long wavelengths is not affected largely
and thus indicates the alignment characteristics of the dust grains directly.  The
information such as the direction of the magnetic field may be most correctly
traced by polarization observations for wavelengths longer than 100\,$\mu$m, though 
spectral observations should be quite important 
to investigate the effects other than the alignment.

On the other
hand the polarization at short 
wavelengths should contain the information of the dust
properties.  
It is generally difficult to separate the effects of
the grain shape and alignment degree solely
from the polarization observation of continuum emission.  
However, the effect of radiation anisotropy may
provide a possible method to separate these factors.
If we know the radiation geometry (location of the heating source)
quite well by other observations, the effect can be estimated to some accuracy.
As shown in Figures~4, 5, and 6, the polarization degree differs at short wavelengths
for grains with different shapes even if the degree at long wavelengths is the same.
A larger polarization degree is expected for flatter oblate grains at short
wavelengths and the differences in the degree among the grains of
$b/a=1.5, 2$, and 3 will be in a distinguishable level ($\sim 1$\%).
If the alignment and radiation directions are not parallel, the change of the
polarization angle should provide an independent confirmation of the derived parameters.
Hence, the effect of anisotropic radiation field
may provide a new tool to investigate the grain shape other than the dust
band \citep{lee85, ait88, 
hen93, hil95}, though in an actual application one may also have to take
account of the effects of dust populations and/or the optical
depth \citep{hil99}.  

The effect of particle size may also have to be examined.
It is coupled with the spectrum of the incident radiation field (Paper I)
and has to be investigated with a radiation spectrum appropriate for the object
in question.
For the ISRF spectrum, the effect of particle size is not significant between 
$a=0.1$\,$\mu$m and 0.03\,$\mu$m because its energy peaks at a
sufficiently long wavelength.  However, in the case where high energy photon
dominates, smaller grains can produce larger polarization (Paper I).

The present study demonstrates the importance of the spectral observations
of polarization in the far-infrared.  Correct information on the alignment
or magnetic field may require careful polarization observations over a range of
wavelengths.
Spectral
observations of the polarization degree and angle in the thermal emission should be 
crucial for the investigation of the effect of the anisotropic radiation field.
The present effect may also be useful in the identification
of the heating source for very obscured objects.



\acknowledgments
This work is stimulated by the ISO polarization workshop held at Vilspa ISO Data Center. 
The author would like to thank the workshop organizers 
for their efforts in the arrangement of the ISO polarization 
workshop.  This work is supported by Grant-in-Aids for Scientific Research 
from Japan Society for the Promotion of Science (No. 10559005 and 11440062).

\clearpage



\end{document}